\documentclass[aps,prl,twocolumn,superscriptaddress]{revtex4}

\usepackage{amsmath}
\usepackage{amsfonts}
\usepackage{amssymb}
\usepackage{epsfig}
\usepackage{graphicx}

\newcommand{\ra}{\rangle}
\newcommand{\la}{\langle}
\newcommand{\be}{\begin{equation}}
\newcommand{\ee}{\end{equation}}

\begin{document}

\title{Large Fluctuations in Driven Dissipative Media}

\author{A. Prados}
\affiliation{F\'{\i}sica Te\'orica, Universidad de Sevilla, Apdo.\ de Correos 1065, Sevilla 41080, Spain}

\author{A. Lasanta}
\affiliation{F\'{\i}sica Te\'orica, Universidad de Sevilla, Apdo.\ de Correos 1065, Sevilla 41080, Spain}

\author{Pablo I. Hurtado}
\affiliation{Instituto Carlos I de F\'{\i}sica Te\'orica y Computacional, Universidad de Granada, Granada 18071, Spain}

\date{\today}

\begin{abstract}
We analyze the fluctuations of the dissipated energy in a simple and general model where dissipation, diffusion and driving are the key ingredients. The full dissipation distribution, which follows from hydrodynamic fluctuation theory, shows non-Gaussian tails and no negative branch, thus violating the fluctuation theorem as expected from the irreversibility of the dynamics. It exhibits simple scaling forms in the weak- and strong-dissipation limits, with large fluctuations favored in the former case but strongly suppressed in the latter. The typical path associated to a given dissipation fluctuation is also analyzed in detail. Our results, confirmed in extensive simulations, strongly support the validity of hydrodynamic fluctuation theory to describe fluctuating behavior in driven dissipative media.
\end{abstract}


\maketitle

Fluctuations are inherent in most physical processes, hinting at the hectic microscopic dynamics that  results in observed macroscopic behavior. They encode essential physical information despite their apparent random origin, and their investigation is opening new paths to understand physics far from equilibrium \cite{LD,Bertini,Derrida,BD,Pablo,GC,LS,iso}. The study of fluctuation statistics of macroscopic observables provides an alternative way to derive thermodynamic potentials, a complementary approach to the usual ensemble description. This observation, valid both in equilibrium and nonequilibrium \cite{Bertini,Derrida}, is most relevant in the latter case where no bottom-up approach exists yet connecting microscopic dynamics with macroscopic properties. The large deviation function (LDF) controlling the statistics of these fluctuations \cite{LD} plays in nonequilibrium physics a role akin to the equilibrium free energy. Key for this emerging paradigm is the identification of the relevant macroscopic observables characterizing nonequilibrium behavior. The system of interest often conserves locally some magnitude (a density of particles, energy, momentum, charge, etc.), and the essential nonequilibrium observable is hence the current or flux the system sustains when subject to, e.g., boundary-induced gradients or external fields. In this way, the understanding of current statistics in terms of microscopic dynamics has become one of the main objectives of nonequilibrium statistical physics, triggering an enormous research effort which has led to some remarkable results \cite{Bertini,Derrida,BD,Pablo,GC,LS,iso}. These recent advances are however restricted to nonequilibrium conservative systems characterized by currents. On the other hand, many nonequilibrium systems are characterized by an irreversible dissipation of energy and need a continuous input of energy in order to stay stationary. Thus the relevant macroscopic observable here is not only the current but also the dissipated energy. This broad class of systems includes granular media, chemical reaction systems, dissipative electronic and biophysical systems, turbulent fluids, and in general all sort of reaction-diffusion systems where dissipation, diffusion and driving are the main ingredients. Fluctuations in dissipative media have been much less studied, most probably because its physics is far more complicated as a result of the irreversibility of their microscopic dynamics.

In this paper we address this issue by studying fluctuations of the dissipated energy in a paradigmatic model of driven dissipative media, using both hydrodynamic fluctuation theory (HFT) and extensive numerical simulations. We consider a generalization of the  Kipnis-Marchioro-Presutti (KMP) model of heat transport \cite{kmp} to dissipative media. The system is defined on a one-dimensional (1D) lattice of size $N$, where each node $i\in [1,N]$ is characterized by an energy $\rho_i$. Dynamics is stochastic and proceeds via collisions between randomly-chosen nearest neighbors $(i,i+1)$ such that a fraction $1-\alpha$ of the total pair energy $\Sigma_i\equiv\rho_i+\rho_{i+1}$ is dissipated out of the system, and the remaining energy $\alpha\Sigma_i$ is randomly redistributed within the pair. Thus, the post-collision energies are $\rho'_i=z\alpha\Sigma_i$ and $\rho'_{i+1}=(1-z)\alpha\Sigma_i$, with $z$ a random number uniformly distributed in $[0,1]$. In addition, boundary sites ($i=1, N$) may interchange energy with a heat bath at temperature $T$ so the system reaches a steady state where energy injection and dissipation balance each other. For conservative dynamics ($\alpha=1$), this model represents at a coarse-grained level the physics of many diffusive systems of technological as well as theoretical interest, and plays a main role in nonequilibrium statistical physics as a benchmark to test theoretical advances \cite{kmp,BD,Pablo,GC,LS,iso}. Our generalization, see also \cite{Levine}, contains the essential ingredients characterizing most dissipative media, namely: (i) diffusive dynamics, (ii) bulk dissipation, and (iii) boundary injection. Moreover, our system can be regarded as a \emph{toy} model for a dense 1D granular gas, in which particles cannot freely move but can collide with their nearest neighbours, losing a fraction of the pair energy and exchanging the rest thereof randomly. The parameter $\alpha$ can be thus considered as the analogue to the restitution coefficient in granular systems \cite{PyL01}. This model is an optimal candidate to study dissipation statistics because: (a) one can obtain explicit predictions for the LDF, and (b) its simple dynamical rules allow for a detailed numerical study. The chances are that our results remain valid for more complex dissipative media described at the mesoscopic level by a similar evolution equation. We report below simulation results for the statistics of the dissipated energy in this model using both standard simulations and an advanced Monte Carlo method \cite{sim}. The latter allows the sampling of the tails of the distribution, and implies simulating a large number $M$ of \emph{clones} of the system.

Our model belongs to a large class of dissipative media characterized at the mesoscale by the following rescaled Langevin equation \cite{PLyH2}
\be
\partial_t \rho(x,t) = -\partial_x j(x,t)-\nu \rho(x,t) \, .
\label{langevin}
\ee
Here $\rho(x,t)$ is the energy density field, with $x\in[-\frac{1}{2},\frac{1}{2}]$ and boundary conditions $\rho(\pm \frac{1}{2},t)=T$, $j(x,t)\equiv Q[\rho(x,t)] + \xi(x,t)$
is the fluctuating current, with local average behavior given by Fourier's law, $Q[\rho]\equiv -D[\rho]\partial_x \rho(x,t)$, $D[\rho]$ is the diffusivity functional, and $\nu$ is the mesoscopic dissipation coefficient. The first term on the rhs of eq. (\ref{langevin}) describes the diffusive spreading of energy, while the second term gives the rate of energy dissipation in the bulk. The current noise term $\xi(x,t)$, which accounts for microscopic random fluctuations at the mesoscopic level, is Gaussian and white with $\la \xi(x,t)\ra=0$ and $\la \xi(x,t)\xi(x',t')\ra=N^{-1} \sigma[\rho] \delta(x-x') \delta(t-t')$, being $\sigma[\rho]$ the mobility functional. This Gaussian fluctuating field is expected to emerge for most situations in the appropriate mesoscopic limit as a result of a central limit theorem: although microscopic interactions can be highly complicated, the ensuing fluctuations of the slow hydrodynamic fields result from the sum of an enormous amount of random events at the microscale which give rise to Gaussian statistics of ${\cal O}(N^{-1/2})$ at the mesoscale.

\begin{figure}
\vspace{-0.5cm}
\centerline{
\includegraphics[width=9cm]{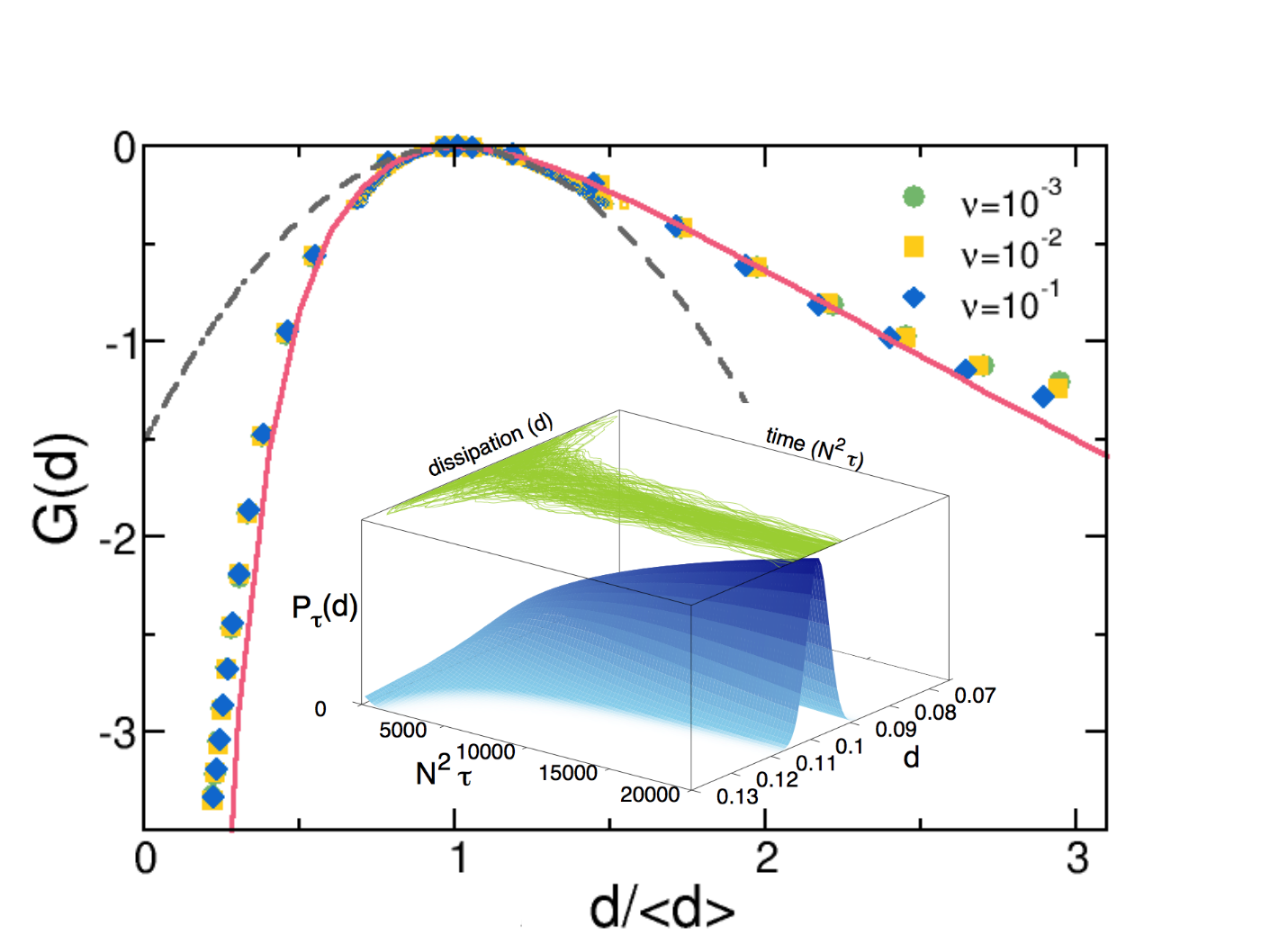}}
\vspace{-0.25cm}
\caption{(Color online) Scaling of the dissipation LDF in the quasi-elastic limit ($\nu\ll1$) for $N=50$, $T=1$ and varying $\nu$. The solid and dashed lines are the HFT prediction and Gaussian estimation, respectively. Small points around the peak were obtained in standard simulations. Inset: Convergence of time-averaged dissipation and sketch of the probability concentration associated to the large deviation principle.}
\label{ldfsmallnu}
\end{figure}

The probability of observing a particular history $\{\rho(x,t),j(x,t)\}_0^{\tau}$ of duration $\tau$ for the density and current fields follows from a path integral and obeys a large deviation principle \cite{LD}, $\text{P}(\{\rho,j\}_0^{\tau}) \sim \exp\{+N\, {\cal I}_{\tau}[\rho,j]\}$, with a rate functional
\be
{\cal I}_{\tau}[\rho,j] = - \int_0^{\tau} dt \int_{-1/2}^{1/2} dx \frac{\displaystyle \Big(j(x,t)+D[\rho]\partial_x\rho(x,t) \Big)^2}{\displaystyle 2\sigma[\rho]} \, ,
\label{HFT2}
\ee
with $\rho(x,t)$ and $j(x,t)$ coupled via the balance equation (\ref{langevin}). Eq. (\ref{HFT2}) expresses the Gaussian nature of local current fluctuations around its average (Fourier's) behavior, and quite remarkably it takes the same form as in many conservative diffusive systems \cite{Bertini,Derrida}. This stems from the subdominant role of the noise affecting the dissipative term in eq. (\ref{langevin}), which scales as $N^{-3/2}$ and thus is negligible against the current noise in the mesoscopic limit \cite{PLyH2}. We are interested in the probability of a fluctuation of the time-averaged dissipated energy $d=\nu \tau^{-1} \int_0^{\tau} dt \int_{-1/2}^{1/2} dx \rho(x,t)$. This probability scales as $\text{P}_{\tau}(d)\sim \exp [+\tau N G(d)]$, see inset to Fig. \ref{ldfsmallnu}, and the dissipation LDF $G(d)$ is related to ${\cal I}_{\tau}[\rho,j]$ via a saddle-point calculation in the long time limit, $G(d)=\tau^{-1} \max_{\rho,j}{\cal I}_{\tau}[\rho,j]$, such that the \emph{optimal} profiles $\rho_0(x,t;d)$ and $j_0(x,t;d)$ solution of this variational problem are consistent with the value of $d$ and the balance equation (\ref{langevin}). These optimal profiles can be interpreted as the ones the system adopts to facilitate a given dissipation fluctuation. For conservative systems it has been shown that optimal profiles equivalent to ours are in fact time-independent in a broad regime as a result of an additivity property of fluctuations \cite{BD,Pablo,iso}, a scenario which may break down only for extreme fluctuations \cite{SSB}. Assuming now time-independent profiles in our dissipative case, the balance equation simplifies to $j'(x)+\nu\rho(x)=0$, with the prime denoting spatial derivative, and
\be
G(d)= - \min_{j(x)} \int_{-1/2}^{1/2} dx \frac{\Big(\nu j(x)-D[-\frac{j'}{\nu}] j''(x)\Big)^2}{2\nu^2 \sigma[-\frac{j'}{\nu}]} \, .
\label{LDF}
\ee
The functional above is a generalized Lagrangian and the associated optimal current profile $j_0(x;d)$ thus obeys a fourth-order Euler-Lagrange equation \cite{PLyH2}, with boundary conditions $j_0'(\pm \frac{1}{2};d)=-\nu T$ and $j_0(-\frac{1}{2};d)=-j_0(\frac{1}{2};d)=d/2$. These follow from the relation $j'=-\nu \rho$, the definition of the dissipation $d$ and the symmetry of our problem around $x=0$.

\begin{figure}
\centerline{
\includegraphics[width=7.5cm,clip]{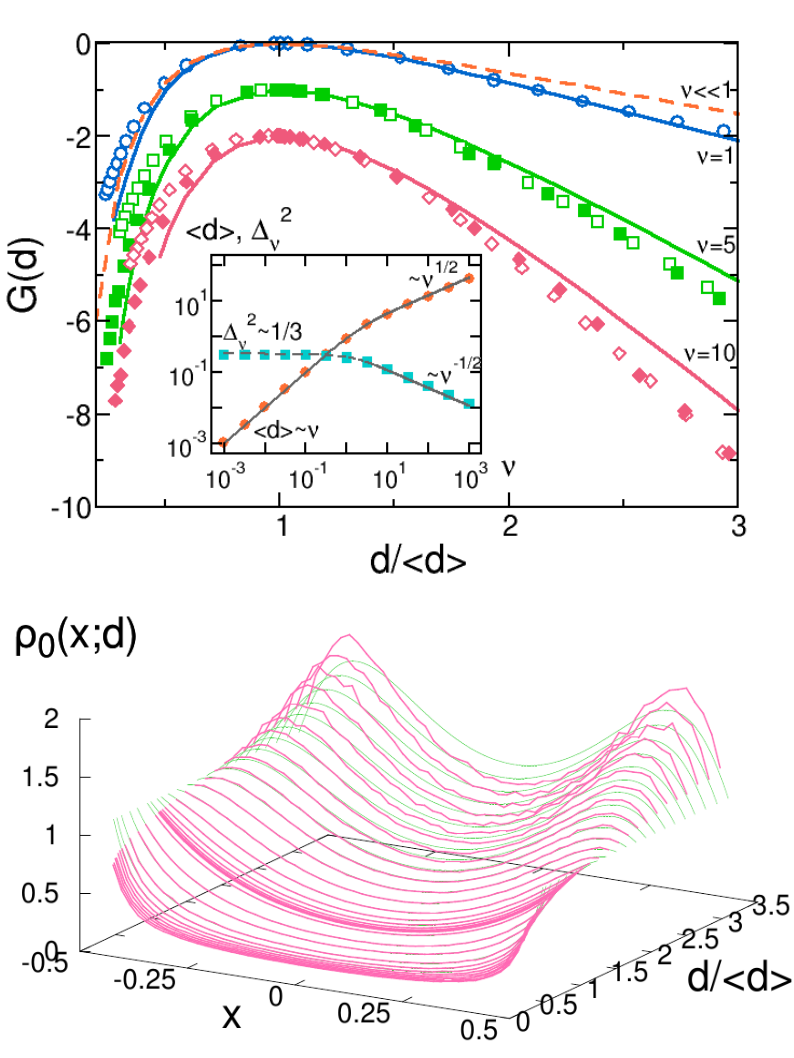}
}
\caption{(Color online) Top: Dissipation LDF measured for increasing values of $\nu$ (shifted vertically for convenience, $G(\la d \ra)=0$ $\forall\nu$). Open symbols correspond to $N=50$ and $M=10^3$ clones, while for filled symbols $N=100$ and $M=10^4$. Solid lines are HFT predictions obtained by numerical integration of eq. (\ref{euler}). Notice the (slow) convergence toward HFT as $N$ and $M$ increase. Inset: Measured average dissipation and its variance as a function of $\nu$, and Gaussian estimation. Bottom: Optimal energy profiles measured for $\nu=10$, $N=50$ and $M=10^3$ (thick lines), and HFT theory.}
\label{ldflargenu}
\end{figure}

For the KMP model $D[\rho]=1/2$ and $\sigma[\rho]=\rho^2$ \cite{kmp,PLyH2}. The stationary average solution to eq. (\ref{langevin}) hence reads
\be
\rho_{\text{st}}(x)=T \frac{\cosh{(\sqrt{2\nu}x)}}{\cosh(\sqrt{\nu/2})} \, ,
 \, j_{\text{st}}(x)=-T \sqrt{\frac{\nu}{2}} \frac{\sinh{(\sqrt{2\nu}x)}}{\cosh(\sqrt{\nu/2})} \, ,
\label{steadysol}
\ee
and the average dissipation is $\la d\ra \equiv \nu\int_{-1/2}^{1/2} dx \rho_{\text{st}}(x)= T \sqrt{2\nu} \tanh (\sqrt{\nu/2})$. Interestingly, these steady profiles are similar to the ones found in vibrated granular gases, when measured in units of the mean-free-path \cite{BRyM00}. The dissipation parameter $\nu$ can be related to KMP microscopic dynamics via $\nu=2(N+1)^2(1-\alpha)$ \cite{PLyH2}. This means that the microscopic dynamics consistent with eq. (\ref{langevin}) must be quasi-elastic ($1-\alpha={\cal O}(\nu N^{-2})$) to guarantee that both diffusion and dissipation interplay on the same scale at the mesoscopic level \cite{Bodineau}. The Euler-Lagrange equation for the optimal current profile solution of the variational problem eq. (\ref{LDF}) for KMP is
\be
12\nu^2j(j'^2-jj'') + j''(3j''-4j'j''') + j'^2j'''' = 0 \, .
\label{euler}
\ee
For small fluctuations of the dissipation, $\epsilon \equiv (d-\la d\ra)/\la d\ra \ll 1$, a regular perturbative analysis of eq. (\ref{euler}) around the steady-state profile eq. (\ref{steadysol}) yields $G(d)\approx -\epsilon^2/2\Lambda_{\nu}^2$, i.e. Gaussian statistics for small fluctuations around the average dissipation, as expected from the central limit theorem. The variance is $\Lambda_{\nu}^2=[\sinh(2\sqrt{2\nu}) - 2\sqrt{2\nu}] / [4 \sqrt{2\nu} \sinh^2(\sqrt{2\nu})]$. Interestingly, $\Lambda_{\nu}^2 \sim 1/3$ independent of $\nu$ for $\nu\ll 1$, anticipating the existence of a simple scaling form for $G(d)$ in this quasi-elastic limit (see below). On the other hand, $\Lambda_{\nu}^2 \sim (2\sqrt{2\nu})^{-1}$ for $\nu\gg 1$, suggesting a suppression of dissipation fluctuations in the strongly inelastic regime. Remarkably, $\Lambda_{\nu}^2$ turns out to be a good estimate for the variance of the dissipated energy, see inset to Fig. \ref{ldflargenu}.
\begin{figure}
\centerline{
\includegraphics[width=9cm]{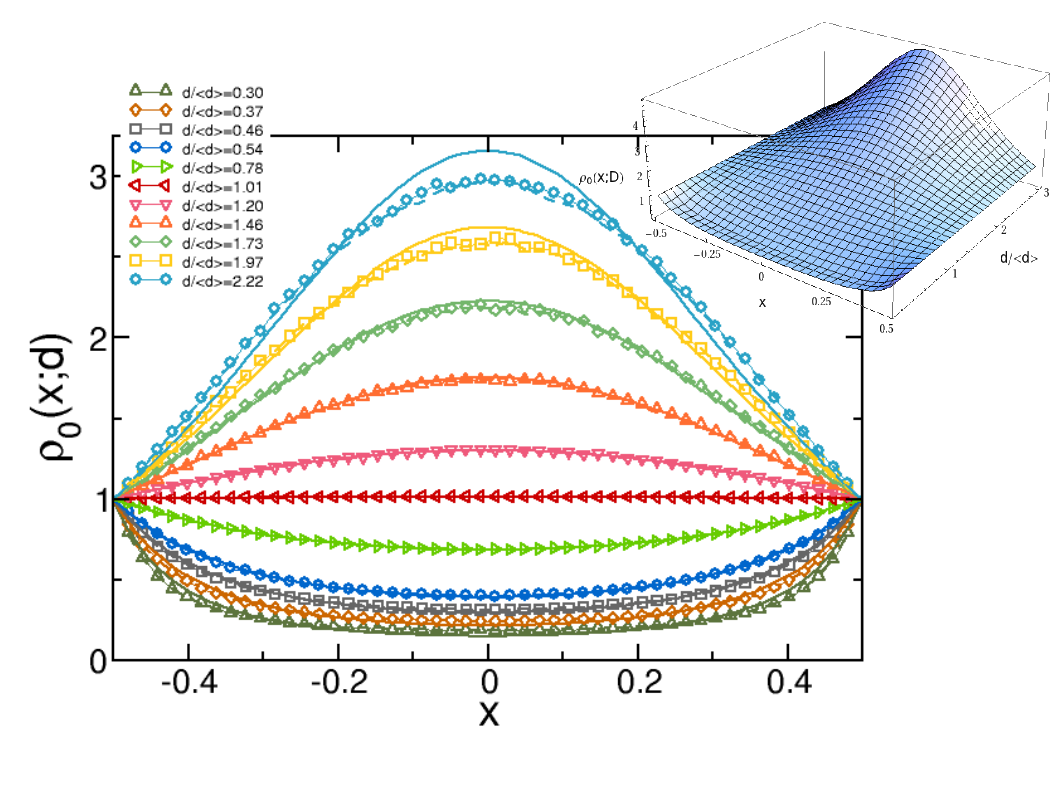}}
\vspace{-0.75cm}
\caption{(Color online) Optimal energy profiles for varying $d/\la d\ra$ measured for $\nu=10^{-3}$ (symbols) and $\nu=10^{-2}$ (dashed lines), and HFT predictions (solid lines and inset).
}
\label{profsmallnu}
\end{figure}

We now obtain the whole dissipation LDF in the quasi-elastic limit $\nu\ll 1$. The behavior of the steady-state current field and average dissipation for $\nu\ll 1$, see eq. (\ref{steadysol}), suggests the scaling $j(x)=\nu \psi(x)$ and $d=\nu\Delta$ in this limit, where $\psi(x)$ and $\Delta$ (the total energy per site) remain of the order of unity. In fact, $\la d\ra\sim \nu T$ for $\nu\ll 1$, i.e. $\Delta\sim T$. Using this scaling in eq. (\ref{euler}) and retaining only the lowest order in $\nu$ we get a differential equation for $\psi(x)$ with solution
\be
\label{17}
\psi_0(x)=-\frac{T}{b} \frac{\tanh\left( bx \right)}{1-\tanh^2\left(\frac{b}{2}\right)} , \quad  \rho_0(x)=T \frac{\cosh^2\left(\frac{b}{2}\right)}{\cosh^2\left(bx\right)} \, ,
\ee
where $\rho_0(x)=-\psi'_0(x)$, and $b$ is a constant given by
\be
\label{18}
\frac{\sinh b}{b}=\frac{d}{\la d\ra}\, . 
\ee
The average behaviour in this quasi-elastic regime is recovered in the limit $b\to 0$ (i.e. $d\to \la d\ra$). For $b\ll  1$ we have small deviations from the average and Gaussian statistics as described above. For arbitrary values of $b$, eq. (\ref{17}) gives the  optimal profiles for the system to sustain an arbitrary fluctuation of the dissipated energy $d$. Using these optimal profiles in eq. (\ref{LDF}) we find for $\nu\ll 1$
\be
G(d) = b\tanh\left(\frac{b}{2}\right) - \frac{b^2}{2},
\ee
with $b$ implicitly given by eq.\ (\ref{18}) in terms of $d/\la d\ra$. This means that, in the low dissipation limit $\nu\ll 1$, $G(d)$ shows a simple scaling form independent of $\nu$ when plotted against the \emph{relative} dissipation $d/\la d\ra$, as anticipated above. The scaling form is fully confirmed in Fig. \ref{ldfsmallnu}, which shows the $G(d)$ as measured for different, small values of $\nu\in [10^{-3},10^{-1}]$ in simulations of the dissipative KMP model. The dissipation LDF is highly skewed with a sharp decrease for fluctuations $d<\la d\ra$ and no negative branch, so fluctuation theorem-type relations linking the probabilities of a given dissipation $d$ and the inverse event $-d$ do not hold \cite{GC,LS}. This was of course expected from the lack of microreversibility, a basic tenet for the fluctuation theorem to apply \cite{Puglisi}. The limit $b\gg 1$ corresponds to large dissipation fluctuations, where $G(d) \approx -\frac{1}{2}[\ln(d/\la d\ra)]^2$, i.e. a very slow decay which shows that such large fluctuations are far more probable than expected within Gaussian statistics ($\sim -\frac{3}{2}(d/\la d\ra)^2$). We also measured the typical energy profile associated to a given dissipation fluctuation, see Fig. \ref{profsmallnu}, finding also very good agreement with HFT. Note that optimal profiles for varying $\nu\ll 1$ also collapse for constant $d/\la d\ra$. Moreover, profiles associated to dissipation fluctuations above the average exhibit an energy overshoot in the bulk. This observation suggests that the mechanism responsible for large dissipation fluctuations consists in a continued over-injection of energy from the boundary bath, which is transported to and stored in the bulk before being dissipated.

For $\nu\gtrsim 1$ we solved numerically eq. (\ref{euler}) and used this optimal profile to obtain $G(d)$. Fig. \ref{ldflargenu} (top) shows the predicted $G(d)$ for increasing, non-perturbative values of $\nu$, and results from simulations. A slow but clear convergence toward the macroscopic HFT prediction is observed as $N$ increases. Such strong finite-size effects are expected since the natural lengthscale associated to a given $\nu$, $\ell_{\nu}=1/\sqrt{2\nu}$, see eq. (\ref{steadysol}), decreases as $\nu$ grows so larger system sizes are needed to observe convergence to the macroscopic limit. In addition, finite-size effects related to the number of clones $M$ used for the sampling become an issue in this limit \cite{Pablo}. In any case, the sharpening of $G(d)$ as $\nu$ increases shows that large dissipation fluctuations are strongly supressed in this regime. In the strongly-inelastic limit, $\nu\gg 1$, the scale $\ell_{\nu}\to 0$ and the system decouples effectively into two independent boundary shells, with energy concentrated around the boundary baths. This is evidenced by the optimal energy profiles for a given $d$ measured for $\nu=10$, see bottom panel in Fig. \ref{ldflargenu}, in contrast to the behavior observed for $\nu\ll1$, see Fig. \ref{profsmallnu}. Moreover, a simple scaling argument shows that $G(d) \stackrel{\nu\gg 1}{=}\sqrt{\nu}{\cal F}(d/\sqrt{\nu})$, with ${\cal F}(z)$ some scaling function \cite{PLyH2}.

Recently a similar hydrodynamic theory has been developed to study large fluctuations in driven dissipative media \cite{Bodineau}, but its predictions do not compare well with numerical results of the dissipative KMP model \cite{PLyH2}. The reason is that Ref. \cite{Bodineau} studies systems with two competing dynamics, one conservative and another non-conservative, thus resulting in independent fluctuations for the density and \emph{dissipation} fields. In our theory, as is the case in many driven dissipative systems, dissipation is linked to the collision process and hence dissipation fluctuations are enslaved to density profile deviations. In fact, both theories coincide in the limit where the optimal dissipation profile is given in terms of the optimal density field  \cite{PLyH2}.

In summary, we have shown that hydrodynamic fluctuation theory is a powerful tool to describe fluctuating behavior in driven dissipative media. This opens the door to further general results in nonequilibrium physics, as HFT provides a well-defined scheme based on the knowledge of the conservation laws governing a system and just two transport coefficients.

\acknowledgments
We acknowledge financial support from Spanish Ministerio de Ciencia e Innovaci\'on projects FIS2008-01339 and FIS2009-08451, EU-FEDER funds, and Junta de Andaluc\'{\i}a projects P07-FQM02725 and P09-FQM4682.

\end{document}